\documentclass[10pt,conference,a4paper]{argencon}

\usepackage[latin1]{inputenc}
\usepackage[compress]{cite}
\usepackage{graphicx}
\usepackage{amssymb}
\begin{document}

% paper title
% can use linebreaks \\ within to get better formatting as desired
\title{A generalized entropy characterization of $N$-dimensional fractal control systems}

\author{\IEEEauthorblockN{Marcos E. Gaudiano\IEEEauthorrefmark{1}\IEEEauthorrefmark{2}\IEEEauthorrefmark{3}$^1$\vspace{0.2cm}}\IEEEauthorblockA{\IEEEauthorrefmark{1}\emph{ATP-group, CMAF, Instituto para a Investiga{\c c}{\~a}o Interdisciplinar. P-1649-003 Lisboa Codex, Portugal.}\\\IEEEauthorrefmark{2}\emph{GRSI-FCEFyN, Universidad Nacional de C\'ordoba. Ciudad Universitaria CP 5000, C\'ordoba, Argentina}\\
\IEEEauthorrefmark{3}\emph{CIEM-CONICET. Ciudad Universitaria CP 5000, C\'ordoba, Argentina.}}\IEEEauthorblockA{$^1$\texttt{\small{gaudiano@famaf.unc.edu.ar}}}}

% conference papers do not typically use \thanks and this command
% is locked out in conference mode. If really needed, such as for
% the acknowledgment of grants, issue a \IEEEoverridecommandlockouts
% after \documentclass

% use for special paper notices
%\IEEEspecialpapernotice{(Invited Paper)}

% make the title area
\maketitle

\begin{abstract}
%\boldmath
It is presented the general properties of $N$-dimensional multi-component or many-particle systems exhibiting self-similar hierarchical structure. 
Assuming there exists an optimal coarse-graining scale $\lambda$ at which the quality and diversity of the (box-counting) fractal dimensions exhibited by a given system are optimized, it is computed the generalized entropy of each hypercube of the partitioned system and shown that its shape is universal, as it also exhibits self-similarity and hence does not depend on the dimensionality $N$. 
For certain systems this shape may also be associated with the large time stationary profile of the fractal density distribution in the absence of external fields (or control). 
\end{abstract}

% no keywords

\IEEEpeerreviewmaketitle

\section{Introduction}
% no \IEEEPARstart
Multi-component, strongly correlated systems, often exhibit non-linear behavior at the microscale leading to emergent phenomena at the macroscale. As P. W. Anderson stated back in 1972, it often happens that 
{\it{"the whole becomes not only more but very different from the sum of its parts"}} \cite{PWA}. Fingerprints of such emergent phenomena can be identified in hierarchical behavior\cite{RavaszBarabasi,AlbertBarabasi} or constraints \cite{PalmerAnderson,CarlosPacheco}, sometimes associated with fractal behavior\cite{Mandelbrot}. 

Hierarchical behavior exhibiting self-similarity has been identified in Physical, Social, Biological and Technological systems\cite{StanleyMeakin}. Employing theoretical tools such as the {\it singularity spectrum} or its equivalent, {\it multiscaling exponents} (via a Legendre's transformation), etc., fractal analysis has been applied to geophysics, medical imaging,
market analysis, voice recognizion, solid state physics, etc. (see e.g. \cite{geophysics, medical,market,voice,solidstate}). Recently, it has been unravelled high quality spatio-temporal fractal behavior\cite{SaraMarcos} in connection with built-up areas in planar embeddings, whose diversity of fractal dimensions covered the entire $[0,2]$ dimensionality spectrum, reflecting the presence of self-organizing principles which strongly constrain the spatial layout of the urban landscape.

Here I investigate in detail the general behavior of multi-component complex systems constrained to exhibit fractal behavior in a space of arbitrary dimensionality $N$. This work is organized as follows. In section \ref{coarse_graining_assumptions} it is defined the central quantity of this work, {\it the entropy $S(D)$ of a cell as a function of its fractal dimension}. In section \ref{some_estimates} there are shown some estimates for $S(D)$ and for the total number of cell fractal configurations. Section \ref{selfsimilar_properties} is about self-similarity properties satisfying $S(D)$ that show that its shape is virtually independent of $N$. Some possible applications of this $N-$dimensional generalization are commented in section \ref{entropically_driven_systems}.

\section{Coarse graining assumptions and the function $S(D)$.}\label{coarse_graining_assumptions}
It will prove convenient to represent our multi-component system in terms of black ($1$) pixels (each a hypercube of side $1$) embedded in a space of dimensionality $N$ otherwise filled with white pixels ($0$). Let us assume the entire system fits 
into a single hypercube of side $\Delta$ (and volume ${\Delta}^N$), and let us divide the system-wide hypercube into a grid of smaller hypercubes (or cells) of side $\lambda$, to each of which it is applied the standard box-counting method ($BCM$) in order to assess the fractal dimension of the system at every cell ({\it location}) \cite{Mandelbrot}.\\
\begin{figure}[h!]
 \centering
\includegraphics[width=0.5\textwidth,height=7cm]{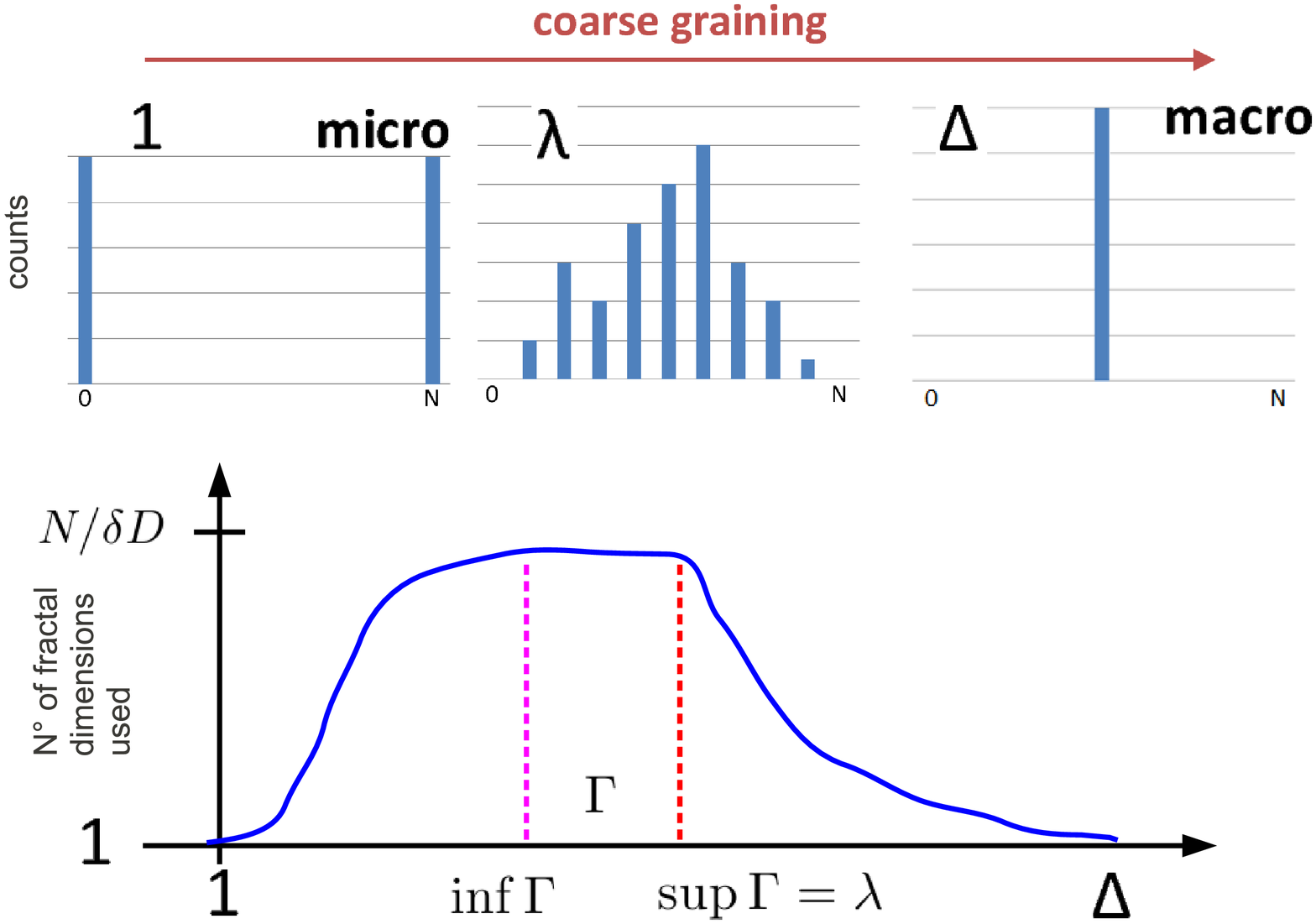}
\caption{Top: The diferent panels illustrate the diversity in fractal dimensions exhibited by the same N-dimensional, hierarchical system, depending on the level of coarse-graining at which the fractal analysis is carried out. Bottom: The same idea, in terms of the number of fractal dimensions used (for a given precision $\delta D$). It is assumed that there exists an intermediate level $\lambda$ at which fractal information is optimal in the sense defined in the main text.}
\label{coarse_graining} 
\end{figure}
Clearly, the number of scales involved in the fractal behavior will depend on $\lambda$ and will always be finite. 
In general, however, there is no reason to expect to find the same fractal dimension at every location, and hence in the following I shall refer to this diversity of fractal dimensions as multi-fractal behavior.\\
The choice of $\lambda$ should not be arbitrary (see Fig.\ref{coarse_graining}). If $\lambda=\Delta$, there will be only one fractal dimension for the entire system. 
On the other hand, if $\lambda=1$ each cell encompasses a single pixel 
resulting in $D=0$ for the (white) pixels that do not belong to the system and, $D=N>0$ for those that do. In between these opposite cases, it is reasonable to assume the existence of some set of coarse-graining levels $\Gamma$ in which the number of fractal dimensions spanned by the $BCM$ is maximized (for a given precision $\delta D$), because the system is multifractal. In addition, the larger the value of $\lambda$, the larger the number of steps that become possible when the $BCM$ is performed, resulting the fractal dimension at every cell to be shared at a larger number of scales and naturally, more properly defined. Thus, we define {\it the optimal coarse-graining level} $\lambda$  ($1<\lambda<\Delta$) as follows:
\begin{equation}
\lambda \doteq \sup \Gamma,
\end{equation}
providing an optimal portray of the multi-fractal behavior of the system.\\

For each cell, the total number of possible configurations is $2^{\lambda^N}$. The volume $V$ of the pixels in each cell that belong to the system can in principle vary between 0 and $\lambda^N$. However, if the system exhibits a local fractal dimension $D$, the number of system configurations is strongly reduced, and it will also depend on the fraction of the cell volume occupied by the system. This conforms with the notion that an emerging property 
of a complex system provides a constraining condition regarding its "entropy" (in a generalized sense), viewed 
here as the log-number of possible configurations of the system compatible with the specified value of the emergent macroscopic variable \cite{Auyang}. 
Before dwelling into the problem of the number of configurations, we shall start by determining lower $L(D)$ and upper $U(D)$ bounds for the volume of a given cell compatible with a pre-defined fractal dimension $D$. 
\par
According to the $BCM$, there exists a minimal number $N_k$ of hyper-boxes of side $2^k$ ($k=0,1,2,...$) 
that cover the pixels of the system inside the cell, which satisfy:
\begin{equation}\label{bcm}
\log N_k = - D k + B
\end{equation}
(throughout this work, $\log$ will be used as a shorthand for $\log_2$) where $D$ is the fractal dimension and $B$ is a constant which is not associated to the fractal behavior\cite{Mandelbrot}. 
In our case, $B$ just shifts the linear fitting up or down, according to the multiple values that $V$ can have. Taking $k=0$ in Eq. (\ref{bcm}) one has that 
$\log N_0=B$, where $N_0$ is the number of pixels that belong to the system, i.e. its volume $V$; thus, $B=\log V$.\\
Recalling that we are dealing with a finite number of scales, we will assume that Eq. (\ref{bcm}) holds for $k=0,...,m$, with $m = [\log (\lambda/2)]$. Hence, in particular we have that $\log V = D m + \log N_m$. On the other hand, $N_m$ is just the number of boxes of edge $2^m$: $1\leq N_m\leq(\lambda/2^m)^N$. Put together, we obtain, 
\begin{equation}\label{LU}
L(D)\doteq 2^{mD}\leq V \leq \lambda^N(2^m)^{D-N}\doteq U(D).
\end{equation}
Thus, there is no configuration with fractal dimension $D$ and with a volume $V$ outside of the bounded region defined by Eq.(\ref{LU}) (the Fig. 1 of \cite{SaraMarcos} is an example of this, for $N=2$). 
It is quite remarkable that the lower bound $L(D)$ is independent of the embedding dimension $N$. 
Knowledge of these constraints on the cell volume compatible with a given fractal dimension $D$ considerably simplifies the determination of the total number of configurations accessible to the system in a cell exhibiting a fractal dimension $D$. 
Let us denote this quantity by $\Omega(V,D)$, where $V$ is the volume occupied by the system in the cell.\\
From Eq. (\ref{bcm}), taking $B=\log V$, the number of boxes covering the part of the system inside
the cell at the $k^{th}$ stage satisfies 
$N_k=N_{k-1} 2^{-D}$. 
$N_k$ is thus defined by $V$ and $D$, but does not depend on the final configuration. In addition, the previous equation 
defines a recursive sequence in which each of the $N_k$ boxes is sub-divided in $2^N$ sub-boxes, $N_{k-1}$ of which will be picked up to enclose the system at the previous $k-1^{th}$ stage. 
The number of ways this procedure can be done is equal to $f(2^N,N_k,N_{k-1})$ (see \cite{setzballs}, \cite{SaraMarcos}), leading to
\begin{equation}\label{OmegaVD}
\Omega(V,D)= \displaystyle{\prod_{k=1}^m f(2^N,V2^{-kD},2^DV2^{-kD})}
\end{equation} 
whereas the total number of configurations having dimension $D$ is given by:
\begin{equation}\label{Omega}
\Omega(D)= \displaystyle{\sum_{V=L(D)}^{U(D)}\Omega(V,D)}.
\end{equation} 
For a given cell of the grid, the function 
\begin{equation}\label{SOmegaD}
S(D)=\log \Omega(D)
\end{equation}
defines a generalized entropy compatible with a cell dimension equal to $D$. For $\lambda=100$ ($m=5$), the function $S(D)$ is the positive one hump function drawn with a {\it blue} thick solid line in Figs. (\ref{SNDeltaNfig}) and (\ref{SdSdD}) (for $N=10$ and $N=70$, respectively). 
\par
Let us conclude this section by remarking something about the function $S(D)$. For another size of the grid, say $\bar{\lambda}$ which, likely, will not correspond to the optimal scale (i.e. $\lambda$), one can similarly define the number of configurations as well as the associated entropy, say $\bar{\Omega}(D)$ and $\bar{S}(D)$. 
Then, if $\lambda$ and $\bar{\lambda}$ do not differ in orders of magnitude (that is, if $m$ remains invariant), one easily finds that: 
\begin{equation}
\Omega(D)\approx \bar{\Omega}(D)^{(\lambda/\bar{\lambda})^N}
\end{equation}
and hence $S(D)$ and $\bar{S}(D)$ will be proportial to each other; consequently, one can say the definition of $S(D)$ is {\it robust to the choice of $\lambda$}.

\section{Some estimates.}\label{some_estimates}
In the following, some estimates for $S(D)$ are computed. These will turn out to be important both to visualize which will be the usual orders of magnitude involved in $\Omega(D)$, as well as to understand some of its main properties.\\
For simplicity, I shall use $U=U(D)$ and $L=L(D)$ except where explicitly indicated. From (\ref{OmegaVD}), (\ref{Omega}) and \cite{fsimplerestimate}, one has that:
\begin{eqnarray}
S(D)&\geq&\log f(2^N,U2^{-D},U)\approx\log{2^{N-D}U\choose U}-1\nonumber\\
&\geq&\lambda^N(2^m)^{D-N}(N-D)-1,
\end{eqnarray}
then, maximizing:
\begin{equation}\label{maxDSD}
\max_D S(D)\gtrsim\frac{\lambda^N}{e\ln 2^m}\geq \frac{\lambda^N}{e\ln \frac{\lambda}{2}}.
\end{equation}
As usual, replacing $\Omega(D)$ by the largest of $\Omega(V,D)$ constitutes a good approximation. Indeed, the relation ${a\choose b}^t \leq {ta\choose tb}$ and \cite{fsimplerestimate} imply that given $D$,  $f(2^N,2^N V/2^{kD},2^DV/2^{kD})$ is an increasing function of $V$ ($k=0,...,m$) as well as $\Omega(V,D)$ (see Eq. (\ref{OmegaVD})); hence, from Eqs. (\ref{OmegaVD}) and (\ref{Omega}), one has that:
\begin{equation}\label{Ssum}
S(D) \leq \log \left( U-L +1\right)+\log \Omega(U,D).
\end{equation}
Then, given that $L\geq 1$:
\begin{equation}\label{SlogOmegaUD}
0\leq S(D)-\log\Omega(U,D)\leq\log U\leq N\log \lambda.
\end{equation}

Thus, from Eq. (\ref{maxDSD}), one can see that (\ref{SlogOmegaUD}) provides a good estimate for $S(D)$, by just assuming
the summation in (\ref{Omega}) to be approximately equal to its greatest summand $\Omega(U,D)$, i.e.:
\begin{equation}
S(D)\approx\log\Omega(U,D)\label{SOmegaUD}.
\end{equation}
The only problem of this estimate occurs at $D=N$: since $\Omega(U(N),N)=1$ one will have from (\ref{SOmegaUD}) that $S(D=N)=0$ which is a excellent estimate, 
despite the fact that $S(D)$ never vanishes. 
Actually, $f(x,y,xy)=1$ implies $\Omega (V,D=N)=\prod_{k=1}^m f(2^N, V/2^{kN}, 2^N V/2^{kN})=1$ for all $V$ so, $S(N)=\log(U(N)-L(N)+1)= \log(\lambda^N-2^{mN}+1)>0$, independently of $N$ (note that $S(N)\approx N\log \lambda$ for large $N$).\\

The equation (\ref{maxDSD}) says that $\Omega(D)$ will usually be a very large number (see Figs. \ref{SNDeltaNfig} and \ref{SdSdD}). For instance, with $N=2$ and $\lambda=100$, it turns out that $\max_D \Omega(D)\approx 10^{1300}$. Nevertheless, the multi-fractal behavior assumed here, together with the existence of an optimal coarse-graining scale $\lambda$ and associated $S(D)$, strongly constrains the number of possible overall configurations the system may explore, dramatically reducing this number compared to an uncorrelated multi-component system:\\ 

{\bf Theorem 1.} 
{\it If our multi-fractal analysis is carried out with a precision -- in dimensionality -- of $\delta D$, then the total number of configurations of a given cell exhibiting a fractal behaviour satisfies
\begin{equation}\label{ntotalfractalconfig}
\sum_{0\leq D\leq N}\Omega(D)\leq N/\delta D\lambda^N2^{\lambda^N a},
\qquad{\textrm{with $a \lesssim 3e^{-\frac{m-1}{m+1}}/2$}.}
\end{equation}}

{\it Proof.}
From (\ref{OmegaVD}), (\ref{SOmegaD}) and (\ref{SlogOmegaUD}) one has that:
\begin{equation}
\Omega(D)\leq \lambda^N \Omega(U,D)=\lambda^N \prod_{k=1}^m f(2^N,U2^{-kD},2^DU2^{-kD}).\label{proof1}
\end{equation}
From \cite{setzballs}, one has that
\begin{equation}
f(2^N,U2^{-kD},2^DU2^{-kD})\leq {2^{N}U2^{-kD}\choose 2^{D}U2^{-kD}}\leq 2^{2^NU2^{-kD}}
\end{equation}
for $k=1,2,...,m$.\\
From (\ref{LU}) and (\ref{proof1}) one obtains:
\begin{eqnarray}
&&\Omega(D)\leq \lambda^N2^{\lambda^N\theta(D)}\label{proof2}\\
&&\theta(D)\doteq(2^{m-1})^{D-N}\frac{1-2^{-mD}}{1-2^{-D}}\leq 3(2^{m-1})^{D-N}/2\nonumber
\end{eqnarray}
for $D\geq D^*=\log(3)\approx 1.585$. 
In the following section, which is independent of this theorem, it is shown that the maximum of $S(D)$ is reached at $D=D_1\approx N-1/\ln(\lambda/2)$. Thus, $D_1>D^*$ for reasonable $N$ and $m$ (i.e. for every $N>2$ and $\lambda\geq 5$, even for $N=2$ and $\lambda\geq 23$). Then, given that $\frac{\lambda}{2}\leq 2^{m+1}$, it follows that:
\begin{equation}
a\doteq\theta(D_1)\leq 3e^{-\frac{m-1}{m+1}}/2.\label{proof3}
\end{equation}
On the other hand,
\begin{equation}
\displaystyle{\sum_{0\leq D\leq N}\Omega(D)}\leq N\Omega(D_1)/\delta D,\label{proof4}
\end{equation}
where $\delta D$ is the precision with which one studies the fractal properties of the system. 
Then, the theorem follows from Eqs. (\ref{proof2}), (\ref{proof3}) and (\ref{proof4}). $\Box$\\

Indeed, the theorem above says that for a given cell, the number of fractal configurations will be just a tiny fraction of the total number of possible, uncorrelated configurations, given by $2^{\lambda^N}$. Let us work out a numerical example. According to (\ref{ntotalfractalconfig}), for $N=2$ and $\lambda=100$ ($m=5$), one finds that $a\lesssim 0.77013$ and
\begin{equation}
\sum_{0\leq D\leq N}\Omega(D)\lesssim 10^{2300}/\delta D,
\end{equation}
which despite being a very big number, it is absolutely negligible with respect to $2^{\lambda^N}\approx 10^{3000}$ (for every reasonable $\delta D$). The theorem explains why almost every random configurations of pixels is not fractal, quantitatively. This result may have applications in Pattern Recognition Theory.

\section{Self-similar properties.}\label{selfsimilar_properties}
The following theorem says that the shape of $S(D)$ is virtually independent of the dimensionality $N$ of the embedding space, actually:\\ 

{\bf Theorem 2.} 
{\it Let us call by $S_N(D)$ and $S_{N+\Delta N}(D)$ to the cell entropy functions (Eq. (\ref{SOmegaD})) for a pair of embedding dimensions $N$ and $N+\Delta N$. The following approximation holds for $D\geq 3$:
\begin{equation}\label{SNDeltaN}
S_N(D)\approx \lambda^{-\Delta N} S_{N+\Delta N}(D+\Delta N).
\end{equation}}

{\it Proof.}
Denoting by $U_N^D$ the upper bound of (\ref{LU}), one has from (\ref{OmegaVD}) and (\ref{SOmegaUD}) that
\begin{equation}\label{SNDlogfk}
S_{N}(D)\approx\sum_{k=1}^m\log f(2^N,U_N^D2^{-kD},2^DU_N^D2^{-kD}). 
\end{equation}
Note that the most important summands of the Eq. above are the ones with $U_N^D2^{-kD}$ and $2^DU_N^D2^{-kD}$ being large numbers ($k=1,2,...$). Then, given that $D\geq 3$, from \cite{fxyzt} and \cite{fsimplerestimate} one can subsequently get to:
\begin{eqnarray}
S_{N}(D)&\approx&2^{-D}\log f(2^N,U_N^D,2^DU_N^D)\label{SNDlogf1}\\
&\approx&2^{-D}\lambda\log f(2^N,U_{N-1}^{D-1},2^DU_{N-1}^{D-1})\nonumber\\
&\approx&2^{-(D-1)}\lambda\log f(2^{N-1},U_{N-1}^{D-1},2^{D-1}U_{N-1}^{D-1})\nonumber\\
&\approx&\lambda S_{N-1}(D-1)
\end{eqnarray} 
(the last equation is obtained by comparison with (\ref{SNDlogf1})). Finally, by iterating the above equation, 
one will obtain Eq. (\ref{SNDeltaN}). $\Box$\\

\begin{figure}[h!]
 \centering
    \includegraphics[width=0.5\textwidth]{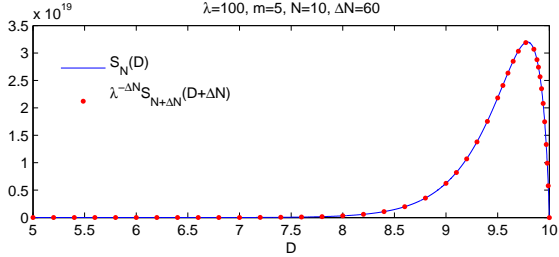}
\caption{Example of application of the self-similar transformation of the Eq. (\ref{SNDeltaN}) for $N=10$, $\Delta N=60$ and $\lambda=100$ ($m=5$).}
\label{SNDeltaNfig}
\end{figure}

\begin{figure}[h!]
 \centering
    \includegraphics[width=0.45\textwidth]{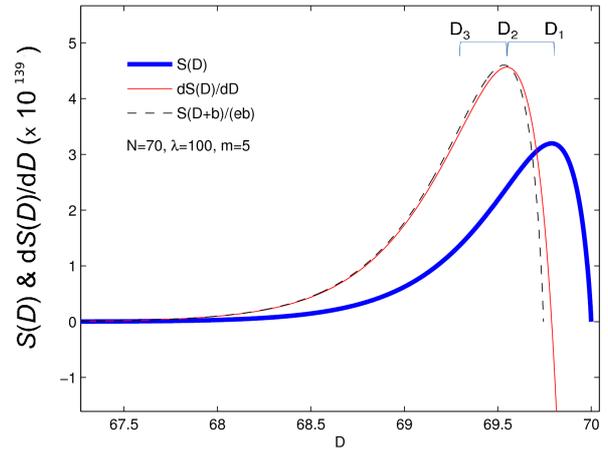}
\caption{Plot of $S(D)$ and $dS/dD$ for $N=70$, $\lambda=100$ ($m=5$). 
The dashed line is the approximation to $dS/dD$ computed with Eq. (\ref{auto-similar}) (see Theorem 3).}
\label{SdSdD}
\end{figure}

%{\it Remark: Note from Eqs. (\ref{SNDlogf1}) that, the approximate Eq. (\ref{SNDeltaN}) converges to a strict equality, exponentially with $D$.}\\
Clearly, the result of the above theorem talks about the existence of self-similarity properties, because the entropies $S_N(D)$ and $S_{N+\Delta N}(D)$ of Eq. (\ref{SNDeltaN}) are related via a translation in $D$ and a multiplication by $\lambda^{-\Delta N}$, which are self-similar transformations (see Fig. (\ref{SNDeltaNfig})).
Something similar occurs between $S(D)$ and its derivative with respect to $D$, showing that the manifestations of self-similarity of the system pervade all quantities (see Fig. \ref{SdSdD}):\\

{\bf Theorem 3.} 
{\it The more orders of magnitude $m=[\log(\lambda/2)]$ are involved in the cell fractal behaviour, the more accurate is the following equation:
\begin{eqnarray}
dS(D)/dD&\approx&S(D+b)/eb\label{auto-similar}\\
\qquad{\textrm{with }}b&\doteq&1/\ln(\lambda/2)\label{b} 
\end{eqnarray}
and $D\leq N-b$.}
 
{\it Proof.} It was mentioned at the end of Section \ref{coarse_graining_assumptions} that the definition of $S(D)$ is robust to choice of $\lambda$.  Thus, let us assume that $m=\log(\lambda/2)$ holds exactly (instead of $m=[\log(\lambda/2)]$).
Evaluating Eq. (\ref{SOmegaUD}) at $D+h\leq N$ ($h\rightarrow 0$), from  \cite{fxyzt} one has that:
\begin{eqnarray}
S(D+h) \approx (\lambda/2)^h\log\Omega(U,D+h)&&\nonumber\\
\approx (\lambda/2)^h\left(S(D)+h\Omega_D(U,D)/\Omega(U,D)\ln 2\right)&&\label{SDh}
\end{eqnarray}
because of Eq. (\ref{SOmegaUD}).
Taking  away $S(D)$ and dividing by $h$, one gets in the limit $h\to 0$
\begin{equation}\label{Sp}
dS(D)/dD\approx S(D)\ln(\lambda/2)+\Omega_D(U,D)/\Omega(U,D)\ln 2.
\end{equation}
Then, eliminating $\Omega_D(U,D)/\Omega(U,D)$ from (\ref{SDh}) and (\ref{Sp}):
\begin{equation}\label{almost}
(2/\lambda)^hS(D+h)-hdS(D)/dD\approx S(D)\left(1-h\ln(\lambda/2)\right).
\end{equation}
Since this equation holds for a range of $h$, one can choose this parameter in order to minimize the quadratic norm of both members, simultaneously. A good ans{\" a}tz for this, is to take 
\begin{equation}\label{hlnl2}
h=1/\ln(\lambda/2)
\end{equation}
which decreases with the number of orders of magnitude involved in the fractal behavior (i.e. $h=(m\ln2)^{-1}$) and it also vanishes the right member of (\ref{almost}). Clearly, the Eqs. (\ref{auto-similar}) and (\ref{b}) follow from (\ref{almost}) and (\ref{hlnl2}). $\Box$\\
 
\begin{figure}[h!]
 \centering
    \includegraphics[width=0.5\textwidth]{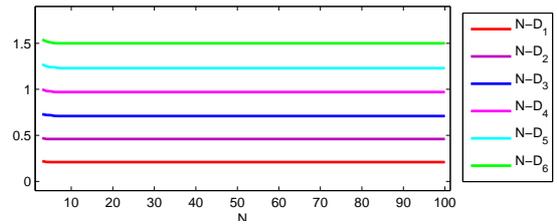}
\caption{Dependence of $D_i$ ($i=1,...,6$) on $N$. For every $N$, the function $S(D)$ was computed for the range $[0,N]$, with a step of $\Delta D=0.01$ and $\lambda=100$ ($m=5$), and its derivatives were approximated by finite differences.}
\label{evenly_spaced}
\end{figure}

A convenient characterization of $S(D)$ is by means of the values $D_i$ satisfying
\begin{equation}
d^i S(D=D_i)/d D^i=0, \qquad{i=0,1,2,...,}
\end{equation} 
Indeed, while $S(D_0)=0$ for $D_0 \approx N$, $D_1$ provides the location of the maximum of the generalized entropy, $D_2$ is associated with the dimension at which entropy reaches its maximum growth rate, etc.. The location of the points $D_i$ is, to a very good approximation, evenly spaced, and one can show by differenciating iteratively Eq. (\ref{auto-similar}) that consecutive $D_i$ satisfy the approximate relation
\begin{equation}\label{Di}
D_i-D_{i+1}\approx b \qquad{i=0,1,2,3,...}
\end{equation}
(see Eq. (\ref{b})). The Fig. (\ref{evenly_spaced}) shows how accurately the Eq. (\ref{Di}) works.
In the multi-fractal system of Ref.\cite{SaraMarcos}, embedded in the spatial dimension $N=2$, the authors have associated the zeros $D_i$ of the derivatives of $S(D)$, with the egdes of different classes or types of urban layouts, exhibiting different levels of risk regarding urban expansion, thus requiring different planning actions from expert policy-makers. Similarly, and given that the shape of the function $S(D)$ is virtually independent of $N$ (Eq. (\ref{SNDeltaN})), the different $D_i$ could play an important role in the characterization of the system.

\section{Entropically driven systems.}\label{entropically_driven_systems}
In Control Theory, it is well-known the fact the controllability of a system depends on the knowledge of the full set of variables that describes the state of the system at a given time. In theory, this principle also applies to Complex Systems but, for almost every system of this kind, there exists a high uncertainty about the numbers of variables that control the system as a whole. In this context, let us assume a kind of {\it very bad possible scenario} in which every part of the system is out-of-control, namely, evolving in time towards a state in which the number of cells with dimension $D$ will be ultimately determined by the entropy $S(D)$ and, having the following density of cells:
\begin{equation}\label{s}
s(D)\doteq\frac{S(D)}{\int_0^NS(x)\,d x}.
\end{equation}
Given that the entropy is a meassure of the ignorance/uncertainty, and assuming that what cannot be controlled is precisely what is ignored, $S(D)$ may provide the degree of uncontrollability of the cells having dimension $D$. Thus, Eq. (\ref{s}) may be seen as a metastable state, result of an hypotetical evolution in the absence of controls (fields/constraints of any kind). In addition, this state may also be recognized as a self-organized one, because $s(D)$ is a non-uniform density and it is assumed that the system can reach it, being driven by its own entropy (purely).\\

%Sciences often attempt to define ideal situations, but for complex control systems it would be very helpful to set the point reference at the worst ideal situation because the system is always doing what we don't want. 
If the system is not dominated by the entropy, the quadratical norm $||\rho-s||_2$ (where $\rho(D)$ is the density of the system) defines how further the system's dynamics is from the one corresponding to the ideal out-of-control worst situation (constituting $s(D)$ as a point of reference). If the system's state is given by (\ref{s}) (or converging to it), one can be sure that the final state will have associated a risky lack of robustness defined as follows. First, it is remarkable the way in which $s(D)$ is concentrated for high dimensions. Actually, independently of $N$, the variance $\sigma^2$ of the fractal dimension reads $\sigma^2\approx 2b^2$ and, the intervals $[D_3,D_1]$ and $[N-1,N]$ approximately incorporate $54\%$ and $76\%$ of the cells, respectively (this follows from the fact the solution of (\ref{auto-similar}) is approximately proportional to $(N-D)e^{(D-N)/b}$). Thus, specially for large $N$, an entropical evolution of the system could ultimately imply a potential lack of robustness and fault-tolerance, in the sense that some external force could affect a big part of the system, despite acting on the cells of a small interval of dimensions.\\ 
On the other hand, it is noteworthy that the rate of growth, as a function of $N$, of the number of possible configurations is by far faster than exponential (see Eq. (\ref{maxDSD})). This means that every time that the embedding dimension increases, the diversity of configurations for fixed $D$ will increase significantly. Thus, assuming that -- which is not necessarily the case -- the density of the system $\rho(D)$ had reached a self-organized equilibrium mimicking $s(D)$ and, if it was possible to lessen the inherent constraints associated with the embedding dimension, increasing the dimensionality from $N$ to $N+\Delta N$, one could let the components of the system to evolve following higher dimensional patterns, opening the possibility for significant diversification. Furthermore, in some cases, even under an entropical evolution, the approach of the system to the new equilibrium given by $S_{N+\Delta N}(D)$ will act to increase the variance $\sigma^2$, improving the robustness of the system, in the sense mentioned above. Indeed:\\

{\bf Theorem 4.} 
{\it Let us denote by $s_N$ and $s_{N+\Delta N}$ to the densities defined by Eq. (\ref{s}) associated to embedding spaces of dimensionality $N$ and $N+\Delta N$, respectively. Let us suppose that the initial cell distribution is given by
\begin{equation}\rho_0(D) = \left\{
\begin{array}{c l}
 s_N(D) \qquad{0\leq D\leq N,}\\
 0 \qquad{N<D\leq N+\Delta N} .
\end{array}
\right.
\end{equation}
Then, if $\rho$ follows the shortest track towards $s_{N+\Delta N}$ , the variance of the fractal dimension will have a maximum in between.}

{\it Proof.} Let us parametrize the evolution of the cell distribution as $\rho_t$, with $0\leq t\leq 1$ ($\rho_0=s_N$ and $\rho_1=s_{N+\Delta N}$). In Functional Analysis, it is shown that 
\begin{equation}\label{shortest}
\rho_t=(1-t)s_N+ts_{N+\Delta N}
\end{equation}
is the shortest track between $s_N$ and $s_{N+\Delta N}$ (independently of the functional norm). In addition, (\ref{shortest}) provides a explicit formula of the variance $\sigma_t^2$ and one can differenciate it twice with respect to $t$, obtaining:
\begin{equation}
d^2\sigma_t^2/dt^2=-2(\Delta N)^2=const.
\end{equation}
(because of Theorem 2). Then:
\begin{equation}\label{sigmatparabola}
\sigma_t^2\approx\sigma_0^2+t(1-t)(\Delta N)^2
\end{equation} 
which is just a parabola having a maximum at $t=1/2$. $\Box$\\

The shortest track of Eq. (\ref{shortest}) is by no means the unique track that makes the variance to increase. Actually, the map $\rho_t\rightarrow \sigma_t^2$ can be considered as a continuous one\cite{explicacion_norma}. This means that, if $\Delta N$ is not excessively small, at least there will exist a positively measurable set of tracks, consisting of paths which are close to the shortest one between $s_{N}$ and $s_{N+\Delta N}$ and consequently, with associated variances $\sigma_t^2$ looking like the one of Eq. (\ref{sigmatparabola}).\\

I conclude by providing some suggestions of problems and systems to which I believe concrete applications of the general ideas developed here could be realized, addressing also the issue of how increasing complexity may evolve in this realm.

Let us consider a given cell of the multifractal system, having fractal dimension $D$ an volume $V$. {\it Per se}, the hyper-pixels of the cell encode all the information about this particular part of the system. 
This encoding is what we may designate by $strategy$ in the sense that it may describe a kind of pattern of behaviour of the part (or agent) of the system associated to the cell (sometimes it could be related to a way to accomplish some goal). For instance,
in urban planning\cite{SaraMarcos}, each strategy encodes a specific pattern of urban layout and land use. In other words, we may associate to each strategy a given (interval of) fractal dimension(s) $D$. If we assume a population of individuals, then different individuals may adopt different varieties of a given strategy, that is, different sequences of hyper-pixels leading to the same $D$. As I have demonstrated here, there is a number of configurations accessible to each $D$, which is given by $S(D)$ and it is constrained by $L(D)$ and $U(D)$. Furthermore it is clear that, at least in some systems \cite{purely_entropical}, and in the absence of control, 
the systems will evolve in time in such a way that the population density of strategies 
${\rho}(D)$ will tend to mimick the normalized $S(D)$. Given the behavior of $S(D)$ mentioned above, this means that the majority of the population will employ strategies lying in a short interval, which corresponds with a low degree of diversity in the number of used strategies.\\
If we think in terms of economics, this means that, country-wide, the natural tendency will be to specialize in connection with a narrow set of activities (say, related to {\it farming} and/or {\it textile}, or to {\it oil production}). In the context of international commerce, this may reflect a tendency for the majority of the commercial actors to find partners or allies in the same area of the world. This, in turn, may be undesirable, and governments and/or agencies may implement policies which constitute the necessary {\it external field} ensuring the systems to remain "{\it far from equilibrium}", with an associated $\rho(D)$ significantly more uniform than $S(D)$ (that is,  diversifying the systems' porfolio \cite{blueoceanstrategy}), being it by fostering the increase of "{\it production means}", being it by increasing the number of "{\it partners}", so as to increase systems' robustness and fault-tolerance. It is also worth mentioning that by increasing the dimensionality of a (hierarchical) system  from $N$ to $N + \Delta N$, one not only increases its inherent complexity but, given the rise of a new available set of strategies, it also paves way for a potential diversification. This may be related to the open problem of understanding the major transitions in evolution \cite{major}.\\
All these last ideas are not the main issue of this paper but they could be the topic of future researches, naturally.

\section{Conclusion.}\label{conclusion}
The main assumption of this work was the existence of a coarse graining level that provides an optimal portray of the multi-fractal behavior of the system.
This work tells neither which part of a given real complex control system can be modelled as a multifractal, nor how to define the associated $N-$dimensional space in which it is embedded \cite{on_hindsight}. Even so, hierarchical systems are ubiquitous in the natural world and, given that the hypotheses on the basis of the ideas studied here are so general, it is likely that the multi-component systems we were dealing with may comprise a non-negligible fraction of the hierarchical systems observed, and to which the principles discussed here do apply.

% use section* for acknowledgement
\section*{Acknowledgment}
This research was initially supported by the grant SFRH/BPD/63765/2009 provided by FCT Portugal. I am in debt with Prof. J. M. Pacheco who made my ideas to flourish and go very far away, as well with the rest of the members of ATP-group (U. of Lisbon) for their helpful discussions, specially with S. Encarna{\c c}{\~a}o because without her precious data, I would never have thought about this work.
% references section
% can use a bibliography generated by BibTeX as a .bbl file
% BibTeX documentation can be easily obtained at:
% http://www.ctan.org/tex-archive/biblio/bibtex/contrib/doc/
% The IEEEtran BibTeX style support page is at:

% http://www.michaelshell.org/tex/ieeetran/bibtex/
%\bibliographystyle{IEEEtran}
% argument is your BibTeX string definitions and bibliography database(s)
%\bibliography{IEEEabrv,../bib/paper}
%
% <OR> manually copy in the resultant .bbl file
% set second argument of \begin to the number of references
% (used to reserve space for the reference number labels box)

%\bibliographystyle{IEEEtran}
%\bibliography{IEEEexample}

\begin{thebibliography}{1}
\bibitem{setzballs} The function $f(x,y,z)$ used in Eq. (\ref{OmegaVD}) corresponds to finding out the number of ways of setting $z$ balls into $y$ boxes with $x$ compartments, leaving at least 1 ball per box \cite{SaraMarcos}. Note that $f(x,y,z)\leq{xy\choose z}$.
\bibitem{fsimplerestimate} Actually, the function $f(x,y,z)$ is introduced in \cite{SaraMarcos} in the form of a Riemann's sum, which is approximately equal to $({xy\choose z}+{y\choose z/x})/2$. Thus, given that ${xy\choose z}\geq{y\choose z/x}^x$ for $x=2^N$, easily one can see that it is a good estimate to take $\log f(x,y,z)\approx \log{xy\choose z}-1$.
\bibitem{fxyzt} $\log f(x,ty,tz)\approx t\log f(x,y,z)$ holds for $t>0$ and large $y$, $z$. From the definition of $f$, it is easily seen that it will satisfy:
$$f(x,ty,tz)=\sum_{r_1+\cdots+r_t=tz}\prod_{k=1}^t f(x,y,r_k)$$
for $t\geq 1$.  Applying the Lagrange's multipliers method, one can see that the greatest part of the summation above is reached for $r_k=z$ ($k=1,...,t$). Then one will have that
$$\log f(x,ty,tz)\approx t \log f(x,y,z)$$
as long as $y$ and $z$ remain large numbers (note that this equation will hold even for $0<t<1$, if $tx$ and $ty$ also remain large).
In particular, from (\ref{OmegaVD}) one has also that $\log\Omega(tV,D)\approx t\log\Omega(V,D)$ holds for large $V$ and $t\geq 1$.
\bibitem{explicacion_norma} This will be the case e.g., if one considers that the tracks are differenciable functions of $t$, the norm of every path $\rho_t$ is given by $\max_t\left( \max_D |\rho_t(D)|+\max_D|\partial\rho_t(D)/\partial t| \right)$ and, the norm of every $\sigma_t^2$ is defined as $\max_t |\sigma_t^2|+\max_t|d\sigma_t^2/dt|$.
\bibitem{purely_entropical} These will be systems to which the concept of energy does not apply in a strict sense, in such a way that system's evolution is purely entropical.
\bibitem{on_hindsight} Although, on hindsight, the model of Ref.\cite{SaraMarcos}, besides providing  a proof of principle of how this formulation can be applied to real cases, also offers some clues of how to address multi-component systems from this perspective.
\end{thebibliography}

% that's all folks
\end{document}